\documentclass[conference,a4paper]{IEEEtran}
\usepackage{xcolor}
\usepackage{balance}

\ifCLASSINFOpdf
  \usepackage[pdftex]{graphicx}
\else
  \usepackage[dvips]{graphicx}
\fi

\ifCLASSOPTIONcompsoc
 \usepackage[caption=false,font=normalsize,labelfont=sf,textfont=sf]{subfig}
\else
 \usepackage[caption=false,font=footnotesize]{subfig}
\fi

\begin{document}

\title{Multiport Network Theory for Modeling and Optimizing Reconfigurable Metasurfaces}

\author{\IEEEauthorblockN{Marco Di Renzo\IEEEauthorrefmark{1} and   
Philipp del Hougne\IEEEauthorrefmark{2}   
}                                     
\IEEEauthorblockA{\IEEEauthorrefmark{1}Universit\'e Paris-Saclay, CNRS, CentraleSup\'elec \\ Laboratoire des Signaux et Syst\'emes, 91192 Gif-sur-Yvette, France}
\IEEEauthorblockA{\IEEEauthorrefmark{2}Universit\'e de Rennes, CNRS, IETR -- UMR 6164, F-35000
Rennes, France}
\IEEEauthorblockA{ \emph{marco.di-renzo@universite-paris-saclay.fr} }
}

\maketitle

\begin{abstract}
Multiport network theory (MNT) is a powerful analytical tool for modeling and optimizing complex systems based on circuit models. We present an overview of current research on the application of MNT to the development of electromagnetically consistent models for programmable metasurfaces, with focus on reconfigurable intelligent surfaces for wireless communications.
\end{abstract}

\vskip0.5\baselineskip

\section{Introduction}
Reconfigurable intelligent surface (RIS) is an emerging technology in wireless communications, which turns wireless networks into programmable environments thanks to the wave manipulation capabilities of programmable metasurfaces \cite{9140329}. One of the main open research challenges in wireless communication theory is the development of electromagnetically consistent models for analyzing and optimizing RIS-aided communication networks \cite{9864116}, \cite{9713744}. This need has recently motivated communication theorists to consider the reintroduction of electromagnetics in communication theory, with special focus on surface electromagnetics when RIS is considered \cite{10415512}.

In this context, circuits theory is viewed as a powerful analytical tool for modeling complex communication systems, as it offers a formal theory and efficient methods for analysis and optimization \cite{5446312}. More specifically, multiport network theory (MNT) constitutes a convenient abstraction model as the perfomance metrics of interest can be expressed in a matrix form, which resembles well-known models utilized in signal processing for multiple-input multiple-antenna (MIMO) systems. The general idea of MNT is that each network element, whether it is a transmitter, a receiver, or a scattering object, can be modeled as a multiport circuit whose ports are appropriately loaded. The first application of MNT to RIS-aided networks was reported in  \cite{9319694}. A recent tutorial on MNT applied to RIS-aided networks can be found in \cite{10574199}.

\section{State of the Art}
In this section, we classify available research works on
MNT with focus on RIS-aided wireless communications.

\subsection{Modeling}
As mentioned, the first application of MNT to RIS-aided networks is reported in \cite{9319694}. The proposed approach is based on impedance parameters. A closed-form expression for the mutual impedances is given in \cite{10133771}, by assuming that the RIS elements can be approximated as loaded dipoles. As elaborated in \cite{10453467}, a model based on MNT leads to a non-linear end-to-end response as a function of the optimization variables. In \cite{10195937}, the MNT model is generalized for application to multipath channels, by representing general scattering objects as multiport networks. In \cite{10476858}, the authors extend the work in \cite{9319694}, by considering Chu's theory. The authors of \cite{9514409}, \cite{10551389} and \cite{10501476} propose the use of scattering parameters in lieu of the impedance parameters, and the structural scattering of RIS is discussed in the latter two papers. The application to fully-connected (non-diagonal) impedance and scattering parameters is analyzed in \cite{10418928}, and it is shown that these architectures are beneficial in the presence of electromagnetic mutual coupling. In \cite{delhougne2024physicscompliantdiagonalrepresentationwireless},  the author introduces a physics-compliant diagonal representation for RISs that are parametrized with non-diagonal impedance matrices. In \cite{semmler2024performanceanalysissystemscoupled}, the authors introduce the concept of decoupling network. In \cite{10511924}, the benefits of utilizing the impedance parameters from the physics point of view are discussed. An MNT model for periodic RISs that account for Floquet's modes is introduced in \cite{10501453}. Finally, the authors of \cite{10008536},
\cite{9856592}, \cite{10501073}, \cite{10310003} have introduced a model based on the coupled dipoles formalism, which is expressed in terms of polarizabilities. A comparison between MNT and the coupled dipoles formalism is available  in \cite{delhougne2023risparametrizedrichscatteringenvironmentsphysicscompliant}.

\subsection{Optimization}
Several optimization algorithms based on MNT models have been proposed recently. The first optimization algorithm was reported in \cite{9360851} for single-antenna transmitters and receivers. The approach is based on the Neumann series approximation in order to efficiently handle the inherent non-linearly of the model. The approach was then utilized in several other papers \cite{10195937}, \cite{9525465}, \cite{10666074}, \cite{wijekoon2024electromagneticallyconsistentmodelingoptimizationmutual}, \cite{10666709}. In \cite{10666709}, in particular, the authors propose the first approach to control the structural scattering by design, and they show that the representation in terms of scattering parameters leads to a faster convergence rate. Optimization methods that avoid the Neumann series approximation are reported in \cite{10476915} and \cite{10352994}. The approach in \cite{10352994}, in particular, can be applied to general multiple-antenna systems.

\subsection{Validation Using Full-Wave Simulations}
Given the flexibility of MNT for modeling and optimizing RIS-aided wireless systems, and, in particular, near-field aspects, such as the electromagnetic mutual coupling, a few research works have recently analyzed the accuracy of communication models obtained by using MNT with the aid of electromagnetic simulators \cite{10666074}, \cite{10576064}, \cite{10246382}, \cite{9901291},  \cite{pettanice2024designoptimizationreconfigurableintelligent},
\cite{pettanice2024multiportnetworkmodelingreconfigurable}, \cite{10528316}. In \cite{10666074}, in particular, MNT is shown to be able to model the structural scattering in a consistent manner. In \cite{pettanice2024designoptimizationreconfigurableintelligent} and
\cite{pettanice2024multiportnetworkmodelingreconfigurable}, in addition, an efficient electromagnetic simulator based on the partial element equivalent circuit method is proposed. In \cite{10528316}, it was shown that a compact MNT-model can be calibrated with a single full-wave simulation even in rich-scattering environments involving metallic walls, extended dielectric objects and significant RIS element structural scattering, highlighting that these scattering entities do not have to be explicitly represented by additional ports, if this is not required for design and optimization purposes.

\subsection{Validation Using Measurements and Testbeds}

Besides electromagnetic simulations, there have been experimental efforts to validate physics-compliant models for RIS-parametrized channels. This oftentimes amounts to end-to-end physics-compliant channel estimation. The first report on this problem \cite{sol2023experimentallyrealizedphysicalmodelbasedwave} was in fact not based on MNT but on an operationally equivalent coupled-dipole formalism. The considered radio environment involves unknown rich scattering, but the results in \cite{sol2023experimentallyrealizedphysicalmodelbasedwave} highlight that there is no need to explicitly model the rich scattering as long as its influence on the coupling between the primary entities (antennas and RIS elements) is correctly estimated. Thereby, the number of parameters to be estimated is kept minimal. In \cite{sol2023experimentallyrealizedphysicalmodelbasedwave}, a gradient-descent approach is utilized to estimate the model parameters and it is further highlighted that this is possible based on non-coherent detection, which greatly alleviates the measurement burden. Efforts involving MNT models are reported in \cite{10501200}, \cite{10669140}, \cite{delhougne2024virtualvnaminimalambiguityscattering}, \cite{delhougne2024virtualvna20ambiguityfree}, \cite{9818951}. In \cite{10501200}, the scattering objects in a multipath environment are explicitly modeled in a statistical manner. In a quest to minimize the number of model parameters and their ambiguities, the authors of \cite{delhougne2024virtualvnaminimalambiguityscattering} and  \cite{delhougne2024virtualvna20ambiguityfree} demonstrated experimentally that at least three distinct and known states per RIS element and coupling between neighboring RIS elements are required, reporting closed-form and gradient-descent based approaches. The same methods were shown to be directly applicable to fully-connected RIS in an experimentally grounded case-study in \cite{delhougne2024physicscompliantdiagonalrepresentationwireless}, by leveraging a physics-compliant MNT-based diagonal representation of fully-connected RISs.  Meanwhile, \cite{10669140} and \cite{9818951} considered a free-space radio environment and demonstrated that parameters estimated in simulation are validated experimentally.

\subsection{Applications in Wireless Systems}
Due to the advantageous tradeoff in terms of electromagnetic accuracy and suitability for optimization, MNT has been utilized for various communication problems. In \cite{10440504} and \cite{arXiv.2410.04110}, it is utilized to evaluate the impact of mutual coupling for channel estimation. In \cite{10279095}, \cite{10032536}, \cite{10375247}, it is utilized to analyze the energy efficiency of wireless systems with simultaneous information and power transfer. Other applications include the analysis and optimization of symbiotic radio systems \cite{10606051}, wireless networks on chip \cite{10528316}, machine learning optimization of scattering parameters \cite{peng2024risnetdomainknowledgedrivenneural}, modeling multi-layer transmissive metasurfaces \cite{10530995}, and the optimization of conformal metasurfaces \cite{10636283}.

\balance

\section{Conclusion}
In this paper, we have summarized the state of the art on the application of MNT for modeling and optimizing RIS-aided wireless communications. Current results have highlighted that MNT is a powerful tool for modeling and optimizing smart radio environments, whose accuracy has been validated with the aid of electromagnetic simulations and empirical measurements. For analytical tractability, most works consider loaded dipoles to model the RIS elements. On the other hand, more complex structures for the unit cells are handled by using semi-analytical methods. In light of the promising tradeoff in terms of analytical tractability, suitability for optimization, accuracy, and engineering insights, more efforts are needed to develop MNT-based models that do not rely on electromagnetic simulations and that can be utilized for general reconfigurable metasurfaces integrating communication, sensing and computing in the wave domain \cite{10515204}.

\section*{Acknowledgment}
The work of M. Di Renzo was supported in part by the European Commission through the Horizon Europe project COVER under grant agreement 101086228, the Horizon Europe project UNITE under grant agreement 101129618, and the Horizon Europe project INSTINCT under grant agreement 101139161, as well as by the Agence Nationale de la Recherche (ANR) through the France 2030 project ANR-PEPR Networks of the Future under grant agreements NF-PERSEUS 22-PEFT-0004, NF-YACARI 22-PEFT-0005, NF-SYSTERA 22-PEFT-0006, NF-FOUND 22-PEFT-0010, and by the CHIST-ERA project PASSIONATE under grant agreements CHIST-ERA-22-WAI-04 and ANR-23-CHR4-0003-01. The work of P. del Hougne was supported in part by the Agence Nationale de la Recherche (ANR) through the France 2030 project ANR-PEPR Networks of the Future under grant agreement NF-YACARI 22-PEFT-0005.

\bibliographystyle{IEEEtran}
\bibliography{biblio}

\end{document}